\documentclass[runningheads]{svmult}

\usepackage{makeidx}   
\usepackage{graphicx}  
\usepackage{subeqnar}  
\usepackage{multicol}  
\usepackage{physprbb}  
\makeindex             

\begin{document}
\title*{Local redshift surveys and galaxy evolution}
\toctitle{Local redshift surveys and galaxy evolution}
%
%
\titlerunning{Local redshift surveys and galaxy evolution}
%
\author{Roberto De Propris\inst{1}
\and Matthew Colless\inst{1}
\and Darren Croton\inst{2}}

\authorrunning{De Propris et al.}
%
%
\institute{
Research School of Astronomy and Astrophysics,\\
Australian National University,\\
Weston, ACT, 2611,\\
Australia\\
\and
Max-Plack Institut f\"ur Astrophysik,\\
Garching, D-85740, Germany\\}

\maketitle              

\begin{abstract}

We present observations of galaxy environmental dependencies using 
data from the 2dF Galaxy Redshift Survey. From a combined analysis
of the luminosity function, Butcher-Oemler effect and trends in H$\alpha$
line strengths we find support for a model where galaxy properties
are mainly set by initial conditions at the time of their formation.

\end{abstract}

\section{Introduction}

Local redshift surveys, undertaken to map the three-dimensional distribution
of galaxies and therefore address topics in cosmology, are also useful to
derive a sample of local galaxies and analyze their properties, such as
luminosity functions and star formation rates. Although the current generation
of surveys is too local to be useful for galaxy evolution studies (but see
descriptions of the DEEP2 and VIMOS surveys in the present volume), the 
2dF and SDSS are adequate to obtain an in-depth description of the local
world of galaxies.

\section{The galaxy luminosity function}
This is sometimes regarded as a 0$^{th}$ order statistics, the simplest
characterization of galaxy populations. It is essential to reproduce its
shape as a first step towards a consistent model of galaxy formation. A
review of this topic is presented in Driver \& De Propris (2003). Here we
compare luminosity functions for the field, clusters and as a function of
local density from the 2dF galaxy redshift survey: since these luminosity
functions are derived from the same catalogue of redshifts and photometry,
they should share most of the selection biases and therefore should be
fairly comparable to gain a picture of environmental effects on galaxy
luminosities.

Figure 1 shows a schematic view of the behaviour of the two relevant
parameters of the luminosity function as a function of density, $\delta_8$,
which is the density, for each galaxy, measured within an 8 Mpc sphere.
We see that:

\begin{itemize}

\item $M^*$ becomes brighter in higher density regions

\item $\alpha$ becomes steeper in these regions

\item This is essentially due to evolution in the $M^*$ and $\alpha$ of
early-type galaxies (i.e. spectroscopically quiescent objects)

\item There is little or no evolution among late-type galaxies

\end{itemize} 

\begin{figure}
\begin{center}
\leavevmode
\includegraphics[width=.6\textwidth]{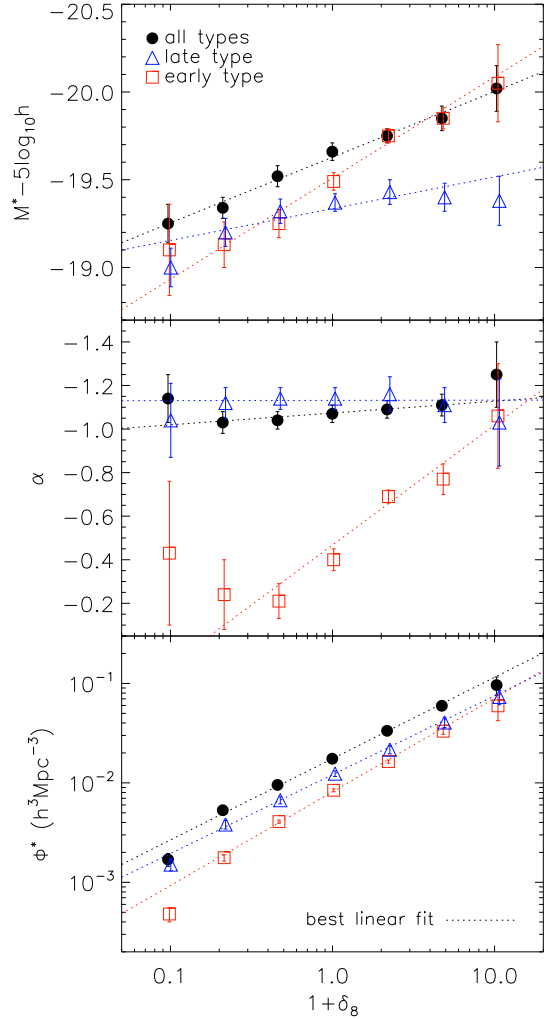}
\end{center}
\caption{The best Schechter fit parameters to the luminosity functions
in the density regimes considered by Croton et al. (2003) for early-type
galaxies (squares) and late-type galaxies (triangles),}
\label{partfig0}
\end{figure}

This is reminiscent of the simple model for galaxy evolution presented
in De Propris et al. (2003), where galaxies were simply assumed to move
from star-forming to quiescent with little luminosity or density evolution.
Although this is a very simplistic scenario, it appears to reproduce the
observations to a good degree. If this is correct, (i) mergers are not
necessary to form galaxies; suppression of star formation suffices and
(ii) even the blue luminosity of most massive galaxies is dominated by
the underlying old population.

\section{The Butcher-Oemler effect}

A traditional measure of star formation and its variation in clusters is
the blue fraction as defined by Butcher \& Oemler (1984). Our sample of
clusters lies at $z < 0.11$ and is therefore unsuitable to study how the
blue fraction evolves, but it allows us to explore how the blue fraction
depends on cluster properties and therefore identify the mechanisms 
responsible, as well as control the selection effects present in cluster
samples at higher redshift.

Figure 2 shows that the blue fraction does not strongly depend on cluster
properties such as richness, velocity dispersion and Bautz-Morgan type.
With proper accounting for errors, the blue fraction may not be significantly
different from cluster to cluster. The only dependencies are with luminosity
(the blue fraction is higher for lower luminosity limits), which argues
that the blue galaxies are intrinsically faint, and with radius.

\begin{figure}
\begin{center}
\leavevmode
\includegraphics[width=.75\textwidth]{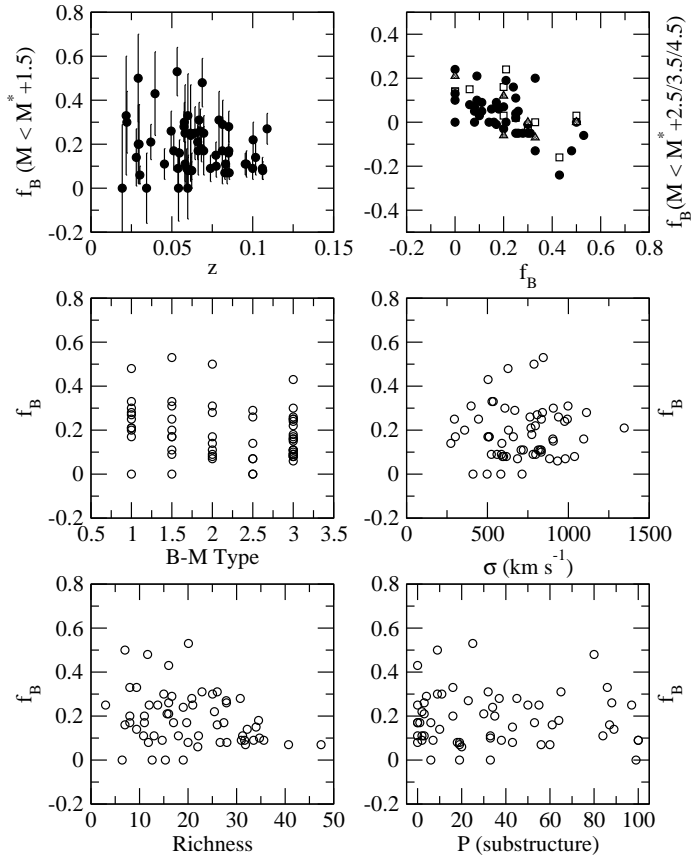}
\end{center}
\caption{Dependence of the blue fraction in clusters on redshift,
luminosity and cluster properties. The top panel on the right shows
the difference between the blue fraction measured to $M^*+2.5$ (filled
circles), $+3.5$ (open squares) and $+4.5$ (triangles) with respect
to the blue fraction measured to $M^*+1.5$ (which is the one used in
all other panels. We omit error bars for clarity, but show them in the
top left panel.}
\label{partfig1}
\end{figure}

The above implies that the original claims for a trend in blue fraction with
redshift may have resulted from a selection effect and an underestimate of
the actual errors and that the cause for the suppression of star formation
responsible for the trend in the luminosity function (and the H$\alpha$
relations shown below) may not be associated with the effects of the
cluster environment (e.g. ram stripping, tides, harassment).

We have explored the Butcher-Oemler effect as a function of the more general
mean density and show that there is a clear relation between lower density 
and higher blue fraction, of which the clusters represent an extreme. Clusters
are simply a continuation of field trends (Figure 3). 

\begin{figure}
\begin{center}
\leavevmode
\includegraphics[width=.675\textwidth]{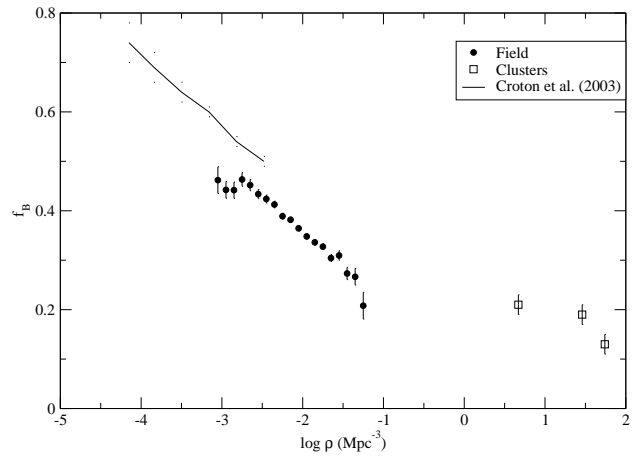}
\end{center}
\caption{Variation of the blue fraction with local density.}
\label{partfig2}
\end{figure}

However, as shown in Figure 4, this is not due to star formation being 
suppressed, but simply to a changing fraction of red and blue galaxies, 
consistent with our simple model for instantaneous suppression of star 
formation with conservation of numbers and luminosity.

\begin{figure}
\begin{center}
\leavevmode
\includegraphics[width=.75\textwidth,angle=-90]{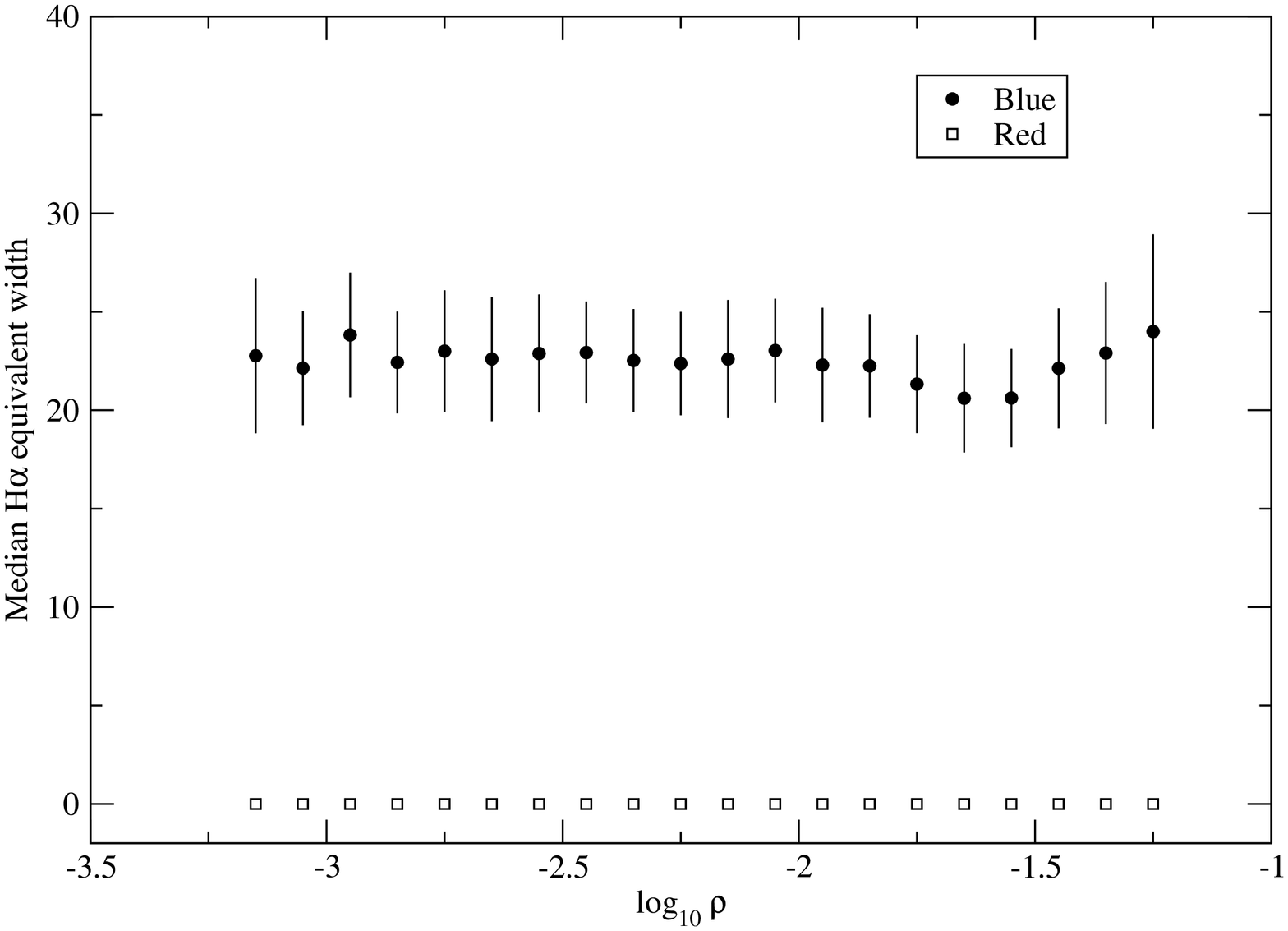}
\end{center}
\caption{Median H$\alpha$ equivalent width for red and blue galaxies as
a function of density.}
\label{partfig3}
\end{figure}

\section{Galaxy ecology: The H$\alpha$-environment relation}

Balogh et al. (2003) have carried this analysis a step further in their study
of trends in H$\alpha$ emission with local density from a combined 2dF and
SDSS sample. Consistent with the previous findings, they show that:
\begin{itemize}

\item The population of galaxies is bimodal: one population shows star 
formation and the other is quiescent

\item The star formation rate is largely independent of density: the numbers
of star forming and quiescent galaxies vary smoothly and continuously with
density.

\item The fraction of star forming galaxies is sensitive primarily to density
over large scales

\end{itemize}

These results argue that the local environment does not influence galaxy
star formation rates, and therefore that galaxy properties are set at high
redshift via processes involving relatively short timescales. This obviously
accounts for the lack of any correlation between the blue fraction and
cluster properties.

\section{A `predestination' model of galaxy evolution}

We present here a speculative scenario for galaxy formation and evolution,
which attempts to account for the observations described above.

In our model, the properties of galaxies are set by their initial conditions.
Galaxies that form in the highest density regions evolve faster and burn
all their gas quickly, therefore becoming ellipticals. Blue galaxies are
a residual population of star-forming objects, which are therefore more
prevalent in lower density regions. The fact that the $U$ (Cortese et al.
2003) and $B$ band luminosity functions for star-forming galaxies are
identical in all environments (De Propris et al. 2003) is a necessary
consequence of this mechanism.

Star formation may be triggered at high redshift in small groups, which
evolve into the dense regions of today. The likely trigger is close
interactions, which have been shown to induce star formation (Barton et
al. 2003), or mergers between gas-rich galaxies. It is unlikely that
these mergers have been important in forming galaxies to $z < 2$, as
also shown by the results of the K20 survey described elsewhere in these
proceedings.

The above scenario is remarkably similar to the original scenario of
galaxy formation in Gaussian random peaks presented by Bardeen et al.
(1986), where the main galaxy properties are set by the initial conditions
at the time of their formation. The slow evolution detected in nearly
all high redshift surveys (with the possible exception of the Combo-17
results presented elsewhere in this volume) is a strong pointer to this
scenario.\\

{\tt The 2dF Galaxy Redshift Survey Team: I. K. Baldry (JHU), Carlton
M. Baugh (Durham), Joss Bland-Hawthorn (AAO), Terry Bridges (AAO),
Russell Cannon (AAO), Matthew Colless$^{PI}$ (RSAA), Chris Collins
(LJMU), Shaun Cole (Durham), Warrick Couch (UNSW), Nicholas Cross (STScI),
Gavin Dalton (Didcot), Roberto De Propris (RSAA), George Efstathiou
(Cambridge), Richard S. Ellis (Caltech), Carlos Frenk (Durham), Karl
Glazebrook (JHU), Edward Hawkins (Nottingham), Carole Jackson (ATNF),
Ofer Lahav (Oxford), Ian Lewis (Oxford), Stuart Lumsden (Leeds), Steve
Maddox$^{PI}$ (Nottingham), Darren Madgwick (UCSC), Peder Norberg (ETHZ), 
John Peacock$^{PI}$ (Edinburgh), Will Percival (Edinburgh), Bruce A.
Peterson (RSAA), Will Sutherland (AAO), Keith Taylor (Caltech)}

%


\begin{thebibliography}{8.}
\addcontentsline{toc}{section}{References}

\bibitem{ba} M. Balogh et al.: Mon. Not. Roy. Ast. Soc. in press (2003)

\bibitem{bbks} J. Bardeen, J. Bond, N. Kaiser, A. Szalay: Astrophys. J.
\textbf{304}, 15 (1986)

\bibitem{bgk} E. Barton Gillespie, M, Geller, S. Kenyon: Astrophys. J.
\textbf{582}, 688 (2003)

\bibitem{bo} H. Butcher, A. Oemler: Astrophys. J. \textbf{285}, 426 (1984)

\bibitem{co} L. Cortese et al.: Astron. Astrophys. \textbf{410}, 25 (2003)

\bibitem{cr} D. Croton et al.: Mon. Not. Roy. Ast. Soc. submitted (2003)

\bibitem{de} R. De Propris et al: Mon. Not. Roy. Ast. Soc. \textbf{342}, 725
(2003)

\bibitem{dd} S. Driver, R. De Propris: Astrophys. Sp. Sci. 
\textbf{285}, 175 (2003)


\end{thebibliography}
\end{document}